\documentclass[aps,twocolumn,prl,superscriptaddress,longbibliography,10pt]{revtex4-1}

\usepackage{graphicx}
\usepackage{amsmath}
\usepackage{amsfonts}
\usepackage{amssymb}
\usepackage{placeins}
\usepackage{helvet}
\usepackage{dcolumn}
\usepackage{nicefrac}
\usepackage{textcomp}
\usepackage{url}
\usepackage{bm}
\usepackage{gensymb}
\usepackage{color}
\usepackage{xspace}
\usepackage{pdfpages}

\newcommand{\cacro}{Ca$_{10}$Cr$_7$O$_{28}$\xspace}
\newcommand{\Cr}{Cr$^{5+}$\xspace}
\newcommand{\musr}{$\mu$SR\xspace}

\begin{document}

\title{Physical realization of a quantum spin liquid based on a novel frustration mechanism}
\author{Christian Balz}
\email[]{christian.balz@helmholtz-berlin.de}
\affiliation{Helmholtz-Zentrum Berlin f\"ur Materialien und Energie, 14109 Berlin, Germany}
\affiliation{Institut f\"ur Festk\"orperphysik, Technische Universit\"at Berlin, 10623 Berlin, Germany}
\author{Bella Lake}
\affiliation{Helmholtz-Zentrum Berlin f\"ur Materialien und Energie, 14109 Berlin, Germany}
\affiliation{Institut f\"ur Festk\"orperphysik, Technische Universit\"at Berlin, 10623 Berlin, Germany}
\author{Johannes Reuther}
\affiliation{Helmholtz-Zentrum Berlin f\"ur Materialien und Energie, 14109 Berlin, Germany}
\affiliation{Dahlem Center for Complex Quantum Systems and Fachbereich Physik, Freie Universit\"at Berlin, 14195 Berlin, Germany}
\author{Hubertus Luetkens}
\affiliation{Laboratory for Muon-Spin Spectroscopy, Paul Scherrer Institut, 5232 Villigen, Switzerland}
\author{Rico Sch\"onemann}
\affiliation{Hochfeld-Magnetlabor Dresden (HLD-EMFL), Helmholtz-Zentrum Dresden-Rossendorf, 01314 Dresden, Germany}
\author{Thomas Herrmannsd\"orfer}
\affiliation{Hochfeld-Magnetlabor Dresden (HLD-EMFL), Helmholtz-Zentrum Dresden-Rossendorf, 01314 Dresden, Germany}
\author{Yogesh Singh}
\affiliation{Indian Institute of Science Education and Research (IISER) Mohali, Knowledge City, Sector 81, Mohali 140306, India}
\author{A.T.M. Nazmul Islam}
\affiliation{Helmholtz-Zentrum Berlin f\"ur Materialien und Energie, 14109 Berlin, Germany}
\author{Elisa M. Wheeler}
\affiliation{Institut Laue-Langevin, 38042 Grenoble, France}
\author{Jose A. Rodriguez-Rivera}
\affiliation{NIST Center for Neutron Research, National Institute of Standards and Technology, 20899 Gaithersburg, USA}
\affiliation{Department of Materials Science, University of Maryland, College Park, 20742 Maryland, USA}
\author{Tatiana Guidi}
\affiliation{ISIS Facility, STFC Rutherford Appleton Laboratory, Oxfordshire OX11 0QX, UK}
\author{Giovanna G. Simeoni}
\affiliation{Heinz Maier-Leibnitz Zentrum, Technische Universitat M\"unchen, 85748 Garching, Germany}
\author{Chris Baines}
\affiliation{Laboratory for Muon-Spin Spectroscopy, Paul Scherrer Institut, 5232 Villigen, Switzerland}
\author{Hanjo Ryll}
\affiliation{Helmholtz-Zentrum Berlin f\"ur Materialien und Energie, 14109 Berlin, Germany}

\date{\today}

\begin{abstract}
\textbf{Unlike conventional magnets where the magnetic moments are partially or completely static in the ground state, in a quantum spin liquid they remain in collective motion down to the lowest temperatures. The importance of this state is that it is coherent and highly entangled without breaking local symmetries. Such phenomena is usually sought in simple lattices where antiferromagnetic interactions and/or anisotropies that favor specific alignments of the magnetic moments are ‘frustrated’ by lattice geometries incompatible with such order e.g. triangular structures. Despite an extensive search among such compounds, experimental realizations remain very few.  Here we describe the investigation of a novel, unexplored magnetic system consisting of strong ferromagnetic and weaker antiferromagnetic isotropic interactions as realized by the compound \cacro. Despite its exotic structure we show both experimentally and theoretically that it displays all the features expected of a quantum spin liquid including coherent spin dynamics in the ground state and the complete absence of static magnetism.}
\end{abstract}

\maketitle

A quantum spin liquid is a macroscopic lattice of interacting magnetic ions with quantum spin number S=\textbf{\textonehalf}, whose ground state has no static long-range magnetic order, instead the magnetic moments fluctuate coherently down to the lowest temperatures \cite{And73,Bal10}. It contrasts with the static long-range magnetically ordered ground states usually observed, and also with spin glass states where the spins are frozen into static short-range ordered configurations \cite{Myd93}. The excitations are believed to be spinons which have fractional quantum spin number S=\textbf{\textonehalf}, and are very different from spin-waves or magnons that possess quantum spin number S=1 and are the characteristic excitations of conventional magnets. Spin liquids exist in one-dimensional magnets and the chain of spin-\textbf{\textonehalf} magnetic ions coupled by nearest-neighbor, Heisenberg (isotropic), antiferromagnetic interactions is a well-established example \cite{Fad81}. This system has no static long-range magnetic order and the excitations are spinons. There is no energetic reason for the spinons to bind together, indeed if a spin-1 excitation is created e.g. by reversing a spin in the chain, it fractionalizes into two spin-\textbf{\textonehalf} spinons \cite{Cau05,Cau05_2,Cau06,Ten95_1,Lak05,Lak13,Mou13,Bal14}.

The existence of spin liquids in dimensions greater than one is much less established. While static order does not occur in one-dimensional magnets, conventional two- and three-dimensional magnets order at temperatures at or above zero Kelvin \cite{Mer66}. This order can be suppressed by introducing competition (known as frustration) between the interactions that couple the magnetic ions. Geometrical frustration is achieved when the magnetic ions are located on lattices constructed from triangular motifs and are coupled by antiferromagnetic interactions.  The antiferromagnetic coupling favors antiparallel spin alignment between nearest neighbor spins which can never be fully satisfied since the number of spins around the triangle is odd. This typically leads to highly degenerate ground states and the tendency for static long-range order is reduced as the system fluctuates among several possible configurations. In quantum spin systems where the magnetic ions have quantum spin number S=\textbf{\textonehalf}, the Heisenberg uncertainty principle produces zero-point motion that is comparable to the size of the spin and which persists down to $T=0$~K. In two- and three-dimensional quantum spin liquids these inherent quantum fluctuations combined with frustrated interactions, suppress static magnetism and give rise to coherent spin dynamics in the ground state \cite{Bal10}.

\begin{figure*}
\includegraphics[width=0.9\textwidth]{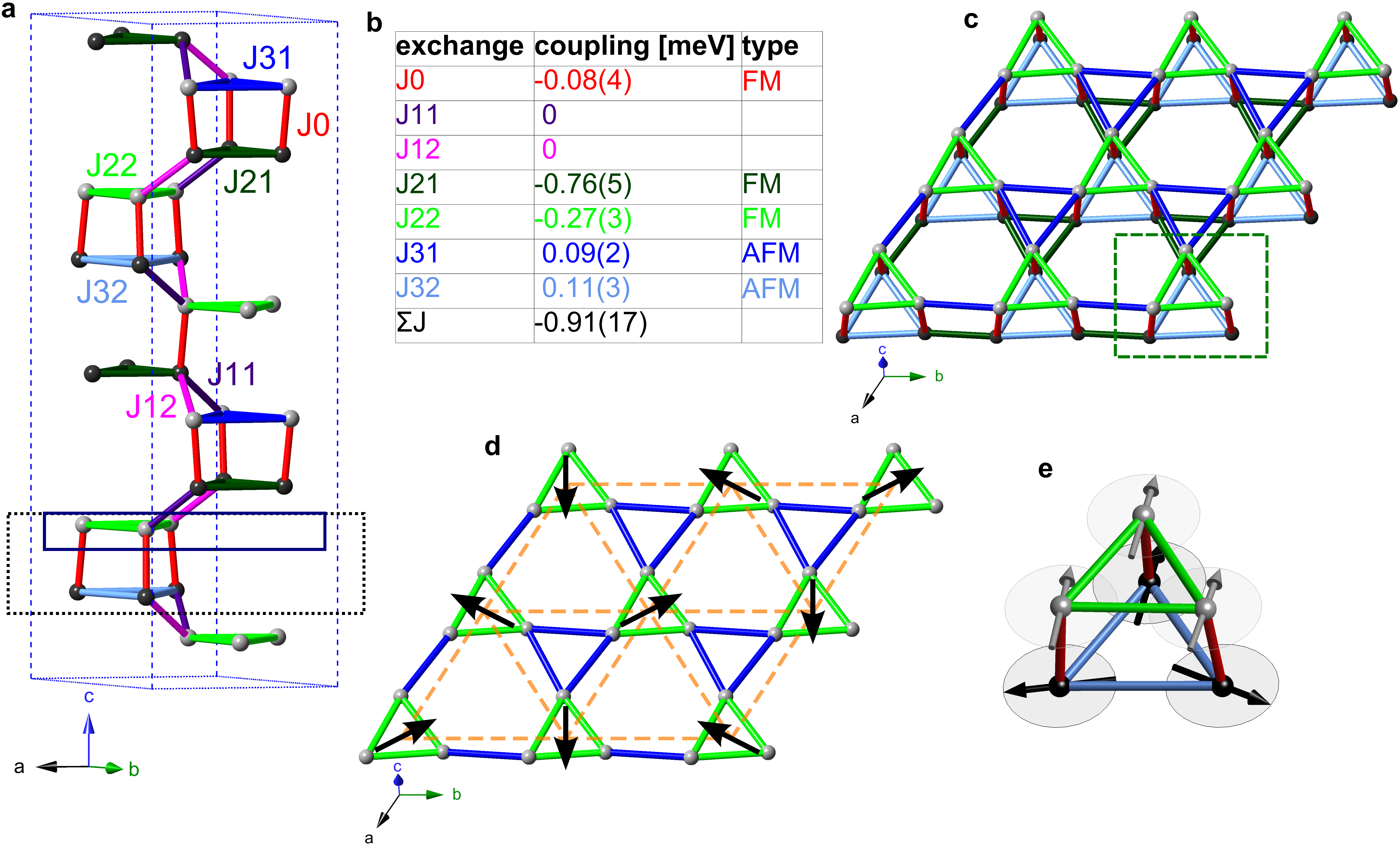}
\caption{\sffamily \textbf{Structure and Hamiltonian.} \textbf{a}, Crystallographic unit cell of \cacro showing only the magnetic \Cr ions which are represented by the black and gray spheres (see supplementary material for details of the crystal structure determination). The seven nearest neighbor couplings between \Cr ions are indicated by the colored lines. The magnetic couplings are isotropic (Heisenberg) exchange interactions and their values are given in (\textbf{b}), FM means ferromagnetic and AFM means antiferromagnetic (see supplementary material for details on how the Hamiltonian and its errorbars were deduced). The crystal structure consists of distorted kagome bilayers of \Cr ions lying in the $ab$ plane as shown in (\textbf{c}). The two layers which form the bilayer both consist of two different corner-sharing equilateral triangles which alternate in the $ab$ plane. The green triangles are ferromagnetically coupled and the blue triangles are antiferromagnetically coupled. It should further be noted that the two layers are inequivalent and have interactions of different sizes. These two layers are coupled together into a bilayer via the ferromagnetic interaction $J0$ (red bond) that connects the ferromagnetic triangles of one layer directly to the antiferromagnetic triangles in the second layer and vice versa. The bilayers were found to be magnetically isolated from each other ($J1=0$) and are stacked along the $c$ axis as shown in (\textbf{a}). \textbf{d}, Kagome single-layer magnetic model. The dashed orange triangles and the black arrows give the 120{\textdegree} order of the effective S=\nicefrac{3}{2} triangular lattice. \textbf{e}, A snapshot of the fluctuating ground state spin arrangement for one bi-triangle consisting of a ferromagnetic upper triangle coupled to an antiferromagnetic lower triangle. This corresponds to the frustrated unit in \cacro.}
\label{fig:structure_hamiltonian}
\end{figure*}

The resonating valence bond state is one of the main conceptual models for a quantum spin liquid \cite{And73}. Here all the spin-\textbf{\textonehalf} magnetic ions form pairs or valence bonds where the two spins have antiparallel alignment and form singlets (total spin zero) resulting in no static order. It arises naturally in the trivial case of the valence bond solid where one dominant antiferromagnetic interaction gives rise to a specific singlet arrangement and gapped magnon (S=1) excitations. The valence bond solid is however not a spin liquid because the fixed spin pairing breaks local symmetry \cite{Rea89}. In contrast, in the resonating valence bond state the valance bonds fluctuate over a number of different configurations with no preference for any particular one leading to long-range entanglement \cite{Lia88}. Excitations can be created e.g. via inelastic neutron scattering by exciting a valence bond from its singlet to its triplet state. Unlike in a conventional magnet this magnon (S=1) excitation can dissociate (or fractionalize) into two spinons (S=\textbf{\textonehalf}) which are able to move apart without costing additional energy, via a simple rearrangement of the valence bonds already superimposed in the ground state \cite{Rea89_2,Rea91,Wen91,Wen02}.

Theoretical work on quantum spin liquids has resulted in number of models consisting of triangular or tetrahedral arrangements of antiferromagnetically-coupled spin-\textbf{\textonehalf} ions. Experimental realizations are in contrast more challenging because real materials usually have additional terms in their Hamiltonians  e.g.\ further neighbor interactions which can lift the degeneracy destroying the spin liquid. The theoretically most promising two-dimensional quantum spin liquid is the kagome lattice consisting of corner-sharing triangles of antiferromagnetically coupled spin-\textbf{\textonehalf} ions \cite{Yan11,Dep12,Wan13,Iqb13,Sut14,Pun14}. Among the many proposed physical realizations of the kagome, the best candidate is the mineral Herbertsmithite which has recently been verified as a quantum spin liquid by the absence of static long-range magnetic order and the presence of spinon excitations \cite{Han12}.

Here we introduce the new quantum spin liquid compound, \cacro. The crystal structure consists of distorted kagome bilayers of \Cr ions (spin-\textbf{\textonehalf}), see figure~\ref{fig:structure_hamiltonian}a \cite{Gye13,Bal16_2}. A single bilayer is shown in figure~\ref{fig:structure_hamiltonian}c, where the inequivalent interactions are represented by the colored lines.  Both the layers forming the bilayer consist of alternating equilateral triangles where the \Cr ions are respectively coupled ferromagnetically (green bonds) and antiferromagnetically (blue bonds). These two layers are coupled together via a weak ferromagnetic interaction (red bonds) connecting the ferromagnetic triangles in one layer directly to the antiferromagnetic triangles in the second layer. The interactions are Heisenberg (isotropic) and their values are given in figure~\ref{fig:structure_hamiltonian}b. This coupling scheme has not been investigated before, and would never be proposed for spin liquid behavior due to the dominant ferromagnetic interactions. Here we prove that \cacro is a quantum spin liquid using a range of experimental techniques. We show that its ground state has no static magnetism and the spins remain dynamically fluctuating down to the lowest temperatures. Furthermore the spectra measured by inelastic neutron scattering are diffuse at all energies as expected for spinon excitations. These experimental findings are confirmed by Functional Renormalization Group calculations.

\begin{figure*}
\includegraphics[width=\textwidth]{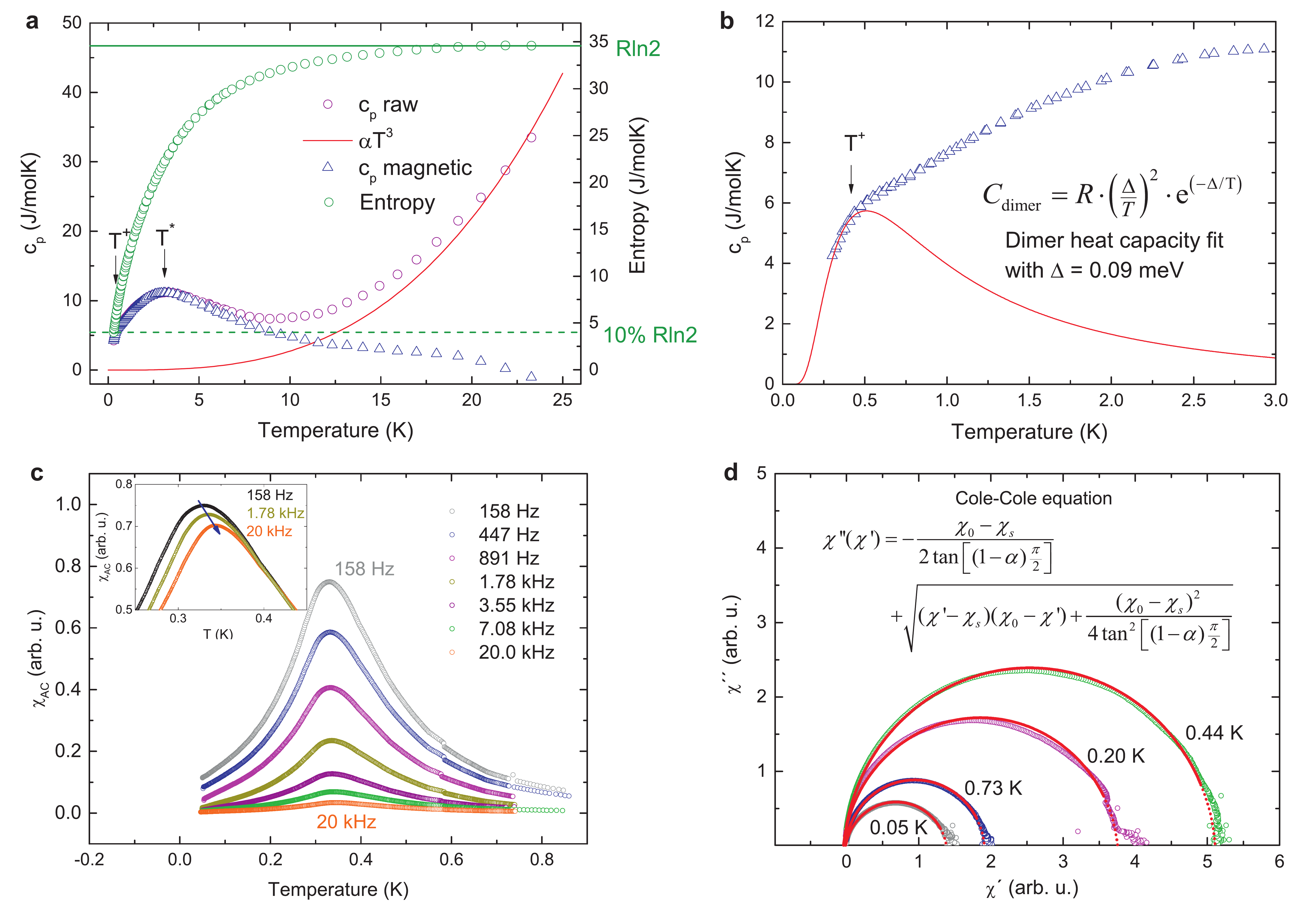}
\caption{\sffamily \textbf{Specific heat and AC susceptibility.} \textbf{a}, Low temperature heat capacity before (purple circles) and after (blue triangles) subtraction of a Debye-like phonon contribution $C_p=\alpha\cdot T^3$ (red line). The magnetic entropy $S=\int\frac{C_p}{T}dT$ where the integration is from 0.3 to 23.3~K (green circles) is plotted on the right axis while the total possible entropy ($R\ln2$ for spin-\textbf{\textonehalf}) is indicated by the green line. \textbf{b}, The magnetic heat capacity below 3~K. The expression for the specific heat of a gapped dimer magnet as shown in the figure is fitted to the data below 0.46~K and allows the upper limit of a possible energy gap to be estimated as $\Delta=0.09$~meV. \textbf{c}, Temperature dependence of the magnitude of the AC susceptibility $\chi_{ac}=\sqrt{\chi'^2+\chi''^2}$ for a range of frequencies $\omega$ measured with $\mu_0H_{\text{dc}}=0$~T and $\mu_0H_{\text{ac}}=30~\mu$T. The scaled frequency-dependent shift of the maximum is shown in the inset. This shift is fitted to $\phi=\frac{1}{T_f}\frac{\Delta T_f}{\Delta\log_{10}\omega}$ where $T_f$ is the temperature corresponding to the observed maximum for the lowest frequency and yields a Mydosh parameter of $\phi=0.022$. \textbf{d}, Cole-Cole plot ($\chi''$ vs. $\chi'$) for different temperatures \cite{Col41}. Each temperature is fitted individually to the Cole-Cole equation given in the figure where data in the frequency range 100 - 20000~Hz was used. The signal below 100~Hz is too weak to obtain reliable data. The Cole-Cole parameter $\alpha$ yields 0.11, 0.06, 0.05 and 0.05 for 50~mK, 200~mK, 440~mK and 730~mK respectively.}
\label{fig:bulk_properties}
\end{figure*}

Among the key features of a quantum spin liquid is the absence of static magnetism in the ground state. A transition to long-range magnetic order can be revealed by a sharp lambda-type anomaly in the temperature-dependent specific heat. The specific heat of \cacro is shown in figure~\ref{fig:bulk_properties}a, no lambda anomaly is evident, proving that no phase transition occurs down to 0.3~K. The magnetic specific heat however shows a broad peak at $T^*\approx3.1$~K indicating the onset of coherent quantum fluctuations and a weak kink at $T^+=0.46$~K suggests a possible crossover at this temperature. The reduced magnetic specific heat $C_p(T)/T$ can be integrated to extract the entropy and reveals that 90\% of the total magnetic entropy is recovered over the temperature range 0.3-23.3~K (figure~\ref{fig:bulk_properties}a). Thus, if a transition occurs below 0.3~K it can release at most 10\% of the entropy, implying that the ordered moment would be highly suppressed.

A common reason for the absence of long-range magnetic order is that the ground state is a valence bond solid, with a static arrangement of dimers. This state has an energy gap which corresponds to promoting a singlet into a triplet. An energy gap between the ground and first excited state is easily observed in the specific heat as an exponential increase at low temperatures. By fitting the data below 0.46~K it can be shown that if a gap exists it must be smaller than 0.09~meV (figure~\ref{fig:bulk_properties}b) thus making a valence bond solid state unlikely.

Another explanation for the absence of a magnetic phase transition is that the ground state is a spin glass. Here the spins are static but have no long-range magnetic order, instead they become locked into short-range ordered configurations below the spin freezing temperature. The freezing temperature is not a true transition and has no signature in specific heat, however a spin-glass can be identified in AC susceptibility by its broad distribution of relaxation times \cite{Myd93}. AC susceptibility measures both the in-phase ($\chi'$), and out-of-phase ($\chi''$) components of the magnetic response of the sample to an alternating magnetic field. By changing the frequency it is possible to probe the relaxation times present in the sample. The magnitude of the AC susceptibility of \cacro as a function of temperature down to 0.05~K for frequencies from 158~Hz to 20~kHz is shown in figure~\ref{fig:bulk_properties}c. A maximum is observed at 0.33~K which is too broad to indicate a phase transition to long-range magnetic order. This peak shifts by 0.015~K over the entire frequency range (Mydosh parameter $\phi=0.022$), such a shift is typical either for an insulating spin-glass or the presence of slow dynamics \cite{Meh15}.

Spin-glass behavior is however ruled out by the Cole-Cole plot where $\chi''$ is plotted against $\chi'$ as the frequency is varied at different fixed temperatures (figure~\ref{fig:bulk_properties}d). A perfect semicircle indicates a single relaxation time whereas a flattened semicircle reveals a distribution of relaxations times \cite{Col41}. The size of the relaxation time distribution is quantified by the Cole-Cole parameter $\alpha$ which takes values from 0 to 1. For \cacro $\alpha$ lies in the range 0.05-0.11 consistent with a single relaxation time, whereas for a typical spin glass it is greater than 0.75 \cite{Myd93}.

\begin{figure}
\includegraphics[width=0.8\columnwidth]{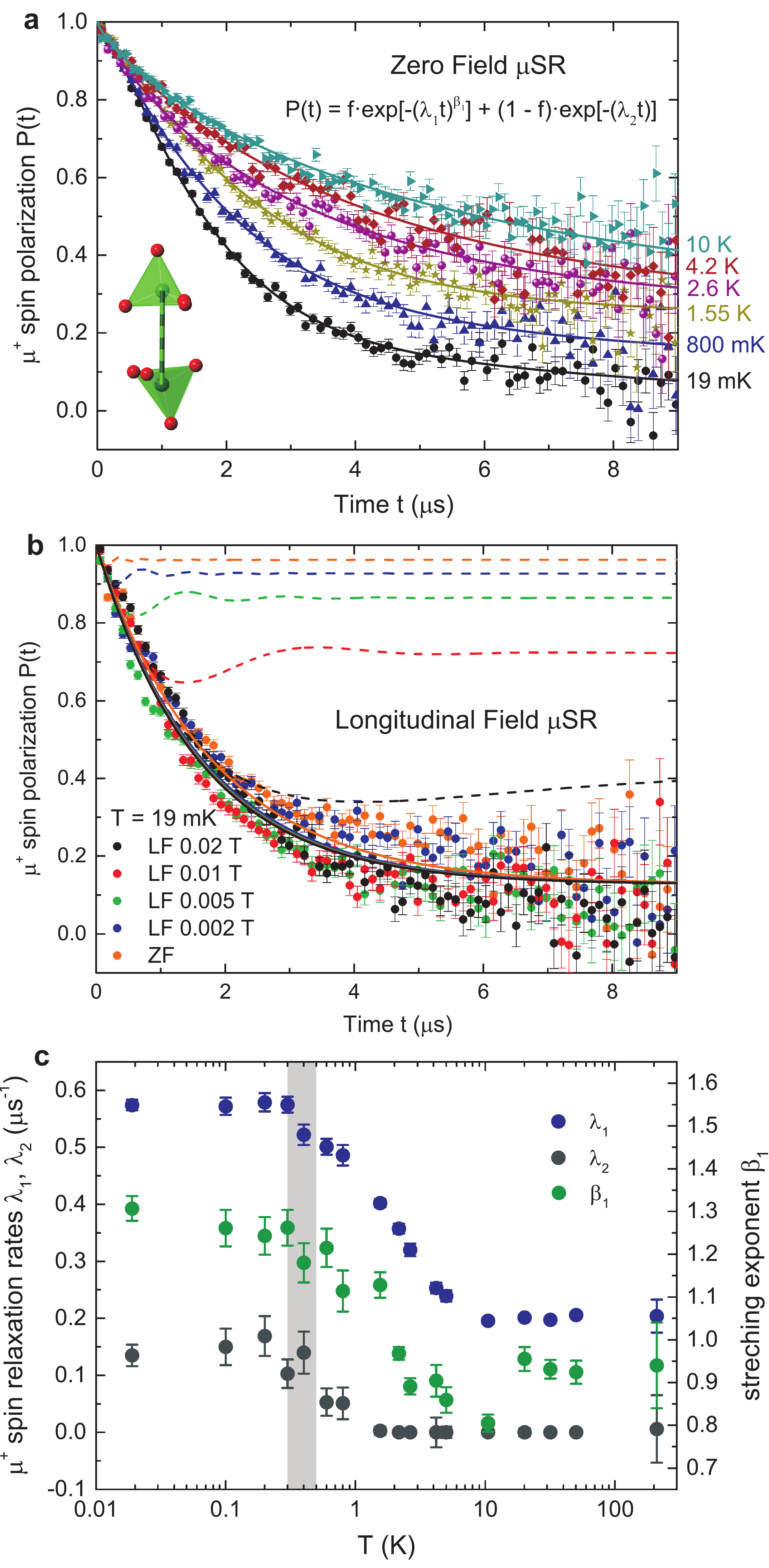}
\caption{\sffamily \textbf{\musr data.} \textbf{a}, Time-dependent muon spin polarization P(t) of \cacro at selected temperatures measured in zero field. Solid lines are fits to the data using the equation shown in the figure. This equation consists of two relaxation processes where $f$ is the fraction of the fast relaxation process, $\lambda_{1}$ and $\lambda_{2}$ are the fast and slow relaxation rates respectively and $\beta_1$ is the stretching exponent of the fast relaxation rate. The two relaxation processes were found to have the weighting 3:1. This is consistent with the ratio of the two muons sites expected from the crystal structure. The inset shows the nearest-neighbor CrO$_4$ tetrahedra, the muons can implant close to either the three in-plane oxygen (giving the fast relaxation) or the one apical oxygen (giving the slow relaxation). \textbf{b}, Data collected in a longitudinal magnetic field at 19~mK. The zero-field data is fitted to the theoretical Kubo-Toyabe functions for both static (dashed black line) and dynamic (solid black line) relaxation \cite{Kub67}. Using the fitted parameters the static (dashed lines) and dynamic (solid lines) relaxation is then simulated for various longitudinal fields. Comparison to the data clearly gives much better agreement with the predictions for a dynamical ground state. The color code of the simulations corresponds to the color code of the data. \textbf{c}, The temperature dependence of $\lambda_1$, $\lambda_2$ and of $\beta_1$ as given in the equation shown in (a). The shaded region from 0.3-0.5~K where the relaxation becomes constant coincides with the temperature where the AC susceptibility goes through a maximum and the specific heat shows a kink. The broad peak at 3.1~K in the specific heat corresponds to the temperature where the relaxation rates start to increase. The errorbars in (a) and (b) represent one standard deviation, the errorbars in (c) are the standard errors of the respective fit parameter.}
\label{fig:muSR}
\end{figure}

Muon spin rotation/relaxation (\musr) is an extremely sensitive local probe of long-range magnetic order able to detect tiny magnetic moments down to 10$^{-4}~\mu_B$ and distinguish between static and dynamic behavior. 100\% spin polarized muons are implanted in the sample at a specific interstitial lattice site and undergo Larmor precession due the local magnetic field at the muon site while. The muon spin polarization as a function of time, P(t), probes the magnetism in the sample. In the case of long-range magnetic order, the magnetic field experienced by the muons is static and P(t) oscillates due to the Larmor precession of the muon spin. In contrast dynamical fields which fluctuate during the muon lifetime rapidly relax P(t) and no oscillations are observed \cite{Yao11}. The time-dependent muon spin polarization of \cacro is shown in figure~\ref{fig:muSR}a for several temperatures. The absence of static long-range magnetic order is immediately apparent from the absence of oscillations for all temperatures down to 0.019~K.  It should be mentioned that oscillations would also be absent for a spin glass because although the internal fields are static, muons implanted in different unit cells would experience different fields and the average polarization would rapidly decay due to the disordered ground state.

Evidence against spin glass freezing comes from \musr measurements in a longitudinal magnetic field. In the presence of static magnetism, a magnetic field applied parallel to the initial muon spin direction shifts P(t) to higher values. At fields $\sim$10 times the internal field the muon spin ensemble maintains its polarization and P(t) is essentially constant in time. If the magnetic system were dynamic however this decoupling of the magnetic system would not be possible for small fields \cite{Yao11}. Figure~\ref{fig:muSR}b shows P(t) measured in several longitudinal fields at 0.019~K. No change in the relaxation is observed up to 200~G revealing a dynamical ground state.

In order to learn more about the dynamical behavior of \cacro, the zero field spectra were fitted to extract the relaxation rate as a function of temperature. The fitted expression is shown in Figure~\ref{fig:muSR}a, two relaxation rates were assumed because of the two muon sites expected from the crystal structure. The temperature dependence of the relaxation rates, $\lambda_i$, are plotted in figure~\ref{fig:muSR}c, they increase upon cooling below 3~K and then become constant below $\sim$0.3-0.5~K. This result shows that between 0.3~K and 0.5~K \cacro crosses over to a low temperature regime where the dynamics become temperature independent and persistent. The $\mu$SR data are remarkably similar to the ones obtained on the well studied quantum spin liquid Herbertsmithite \cite{Men07}.

Together heat capacity, AC susceptibility and \musr reveal the complete absence of static magnetism in \cacro. Both static long-range magnetic order and spin-glass freezing are absent down to 0.019~K. They furthermore exclude a spin gap greater 0.09~meV thus making a valence bond solid scenario unlikely. Most importantly \musr and AC susceptibility show that the spins are fluctuating coherently down to the lowest temperatures and that the fluctuation rate becomes constant below 0.3~K revealing a crossover into a regime characterized by persistent slow dynamics as would be expected for a quantum spin liquid.

\begin{figure*}
\includegraphics[width=0.8\textwidth]{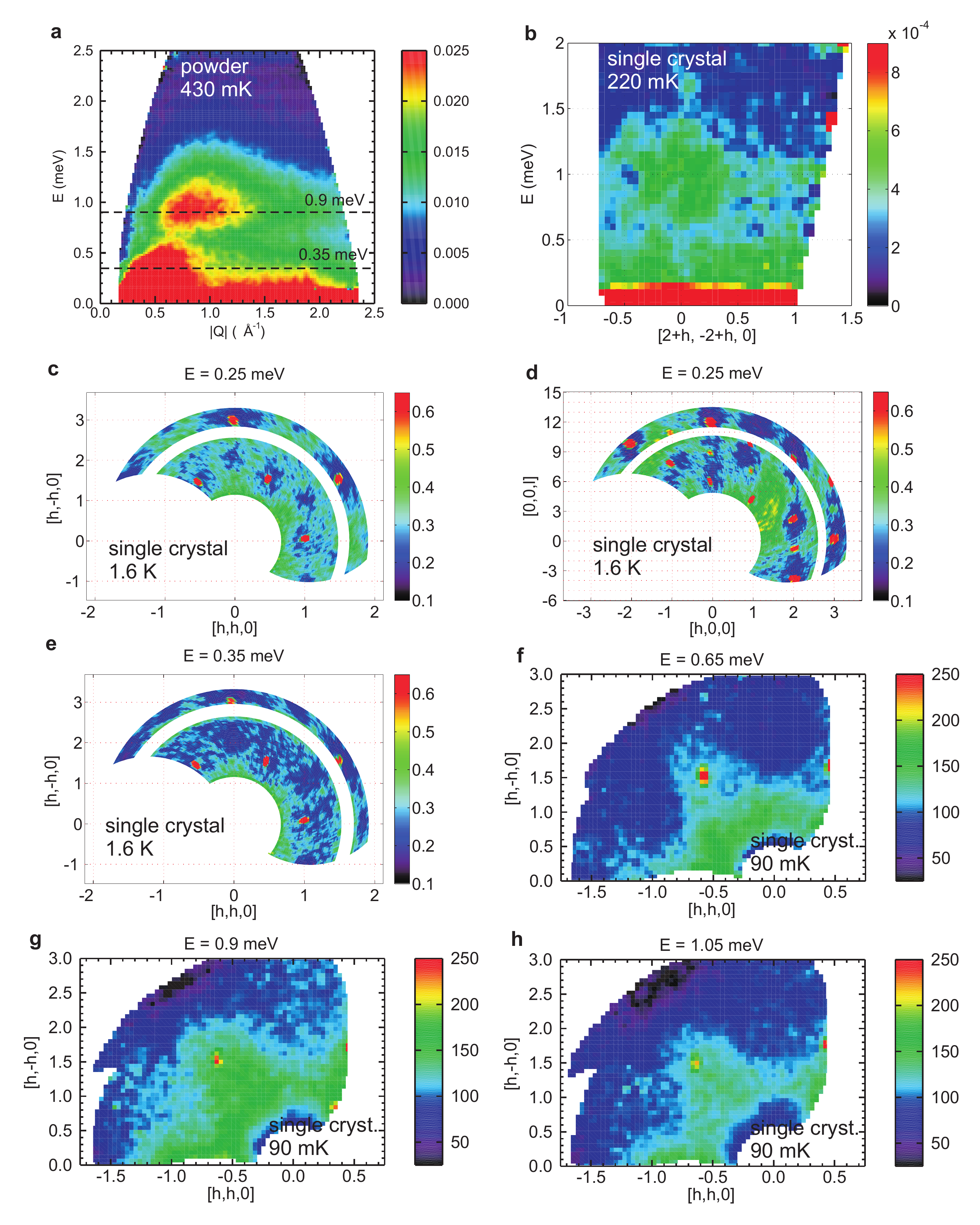}
\caption{\sffamily {\bf Inelastic neutron scattering data measured in zero applied magnetic field.} \textbf{a}, Powder data measured on TOFTOF at $T=430$~mK using an incident energy of $E_i=3.27$~meV. The data are plotted as a function of energy and wavevector transfer. The energy transfers of the constant energy slices shown in (\textbf{e}) and (\textbf{g}) are indicated by the horizontal dashed lines. Data collected using higher incident energies reveals the complete absence of magnetic signal above 1.6~meV. \textbf{b}, Single crystal data measured on OSIRIS at $T=220$~mK. The data are plotted as a function of energy and wavevector transfer parallel to [2+h,-2+h,0] and integrated over the perpendicular wavevector range $1.9<[k,-k,0]<2.1$. There are two bands of excitations at 0-0.6~meV and 0.7-1.5~meV, the low energy signal at 0-0.2~meV is dominated by nuclear and incoherent scattering. Constant energy slices were measured at 1.6~K on a single crystal using the IN14 spectrometer. Data is shown in the plane of the kagome bilayer ($(hk0)$ plane) at \textbf{c} 0.25~meV, and \textbf{e} 0.35~meV, and perpendicular to the bilayers ($(h,0,l)$ plane) at \textbf{d} 0.25~meV. These energies correspond to the lower excitation band. Single crystal data were also collected in the upper excitation band in the $(hk0)$ plane at $T=90$~mK using the MACS II spectrometer at energy transfers of (\textbf{f}) 0.65, (\textbf{g}) 0.9 and (\textbf{h}) 1.05~meV. The red high intensity points in (\textbf{c})-(\textbf{h}) are phonons dispersing from Bragg peaks.}
\label{fig:INS}
\end{figure*}

To explore the dynamics of \cacro in more detail we used inelastic neutron scattering. The neutron scattering cross-section is directly proportional to the dynamical structure factor $S(Q,\omega)$ which is the Fourier transform in space and time of the spin-spin correlation function and allows the magnetic excitations to be mapped out as a function of energy and wavevector transfer \cite{Squ96}. Figure~\ref{fig:INS}a shows the powder spectrum, the excitations appear gapless and form two distinct bands with energy ranges 0.0-0.6~meV and 0.7-1.5~meV. No magnetic scattering is found above 1.6~meV. Measurements on a single crystal were also performed. Figure~\ref{fig:INS}b shows the excitations as a function of energy and wavevector within the plane of the kagome bilayers. While confirming the presence of two bands it is clear that the excitations are dispersionless and much broader than the instrumental resolution of 0.025~meV. Figures~\ref{fig:INS}c \& e-h show the in-plane excitations at various fixed energy transfers. The magnetic scattering is broad and diffuse at all energies. The scattering in the lower band (Fig.~\ref{fig:INS}c \& e) forms diffuse ring-like features very different from the diffuse blocks of scattering observed in the upper band (Fig.~\ref{fig:INS}f-h), however within each band the excitations evolve only gradually with energy. Figure~\ref{fig:INS}d shows the spectrum perpendicular to the kagome bilayers at 0.25~meV, the signal does not disperse along the out-of-plane direction ($[0,0,l]$) proving that the bilayers are magnetically isolated from each other and the magnetism is essentially two-dimensional (further evidence for the two-dimensionality is given by high magnetic field data in the supplemental material).

\begin{figure*}
\includegraphics[width=\textwidth]{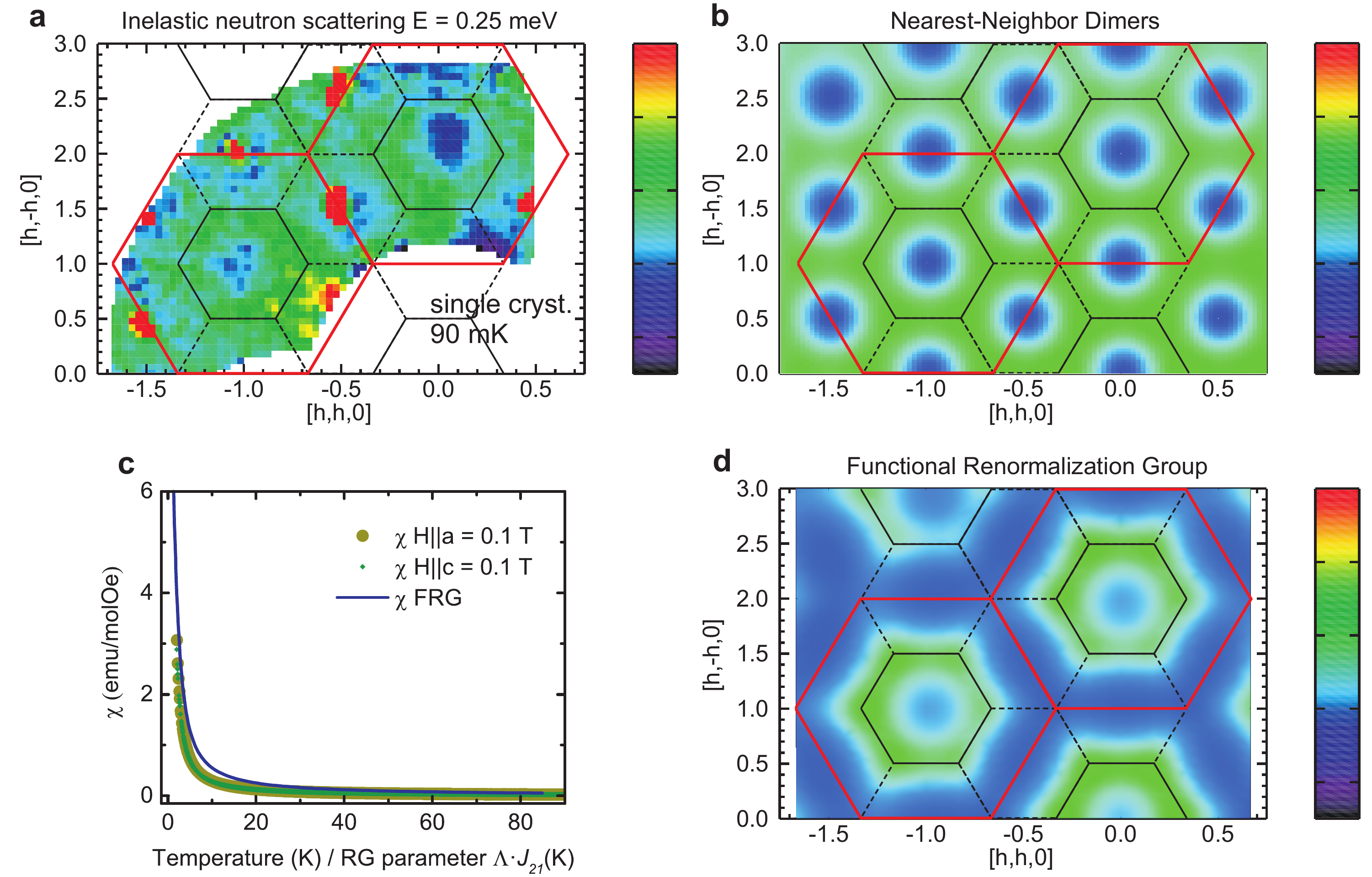}
\caption{\sffamily \textbf{Inelastic neutron scattering data measured in zero applied magnetic field compared to theory.} \textbf{a}, Inelastic neutron scattering intensity in the plane of the kagome bilayers ($(hk0)$ plane) measured on MACS II at $E=0.25$~meV, $T=90$~mK. The black (red) lines are the boundaries of the first (fourth) Brillouin zone. The red high intensity points are phonons dispersing from Bragg peaks. \textbf{b}, Equal-time structure factor for dimers randomly arranged on a triangular lattice using the method described in Ref.~\cite{Han12}. The lattice spacing is assumed to be twice the average in-plane nearest neighbor \Cr-\Cr distance corresponding to the effective S-\nicefrac{3}{2} triangular lattice proposed for \cacro (dashed orange lines in Fig.~\ref{fig:structure_hamiltonian}d). The calculated intensity was multiplied by the magnetic form factor of \Cr. This model captures the ring-like intensity modulation of the data. \textbf{c},~DC susceptibility measured with a magnetic field of $H=0.1$~T applied parallel to the $a$ and $c$ axes plotted alongside the PFFRG susceptibility at $Q=0$ calculated for the Hamiltonian of \cacro (see supplementary material for further details). \textbf{d}, $Q$-space resolved susceptibility at $E=0$~meV and $T=0$~K calculated via PFFRG. The signal is corrected for the distortion of the lattice from ideal kagome symmetry as well as for the magnetic form factor of \Cr. Diffuse hexagonal-shaped scattering is obtained with broad maxima at the corners of the first Brillouin zone.}
\label{fig:INS_theory}
\end{figure*}

The presence of broad, diffuse excitations contrasts strongly to the sharp and dispersive spin-wave modes typical of conventional magnets with long-range magnetically ordered ground states. Diffuse excitations are also incompatible with a valence bond solid where a fixed arrangement of singlets in the ground state gives rise to well-defined gapped modes. Valence bond solid order is probably prevented because the kagome bilayers of \cacro cannot be patterned by nearest-neighbor dimers without putting a dimer on a ferromagnetic bond resulting in a large energy cost. On the other hand the excitations of a quantum spin liquid are expected to be spinons which have a spin quantum number of S=\textbf{\textonehalf}. In inelastic neutron scattering, spinons are observed as a broad continuous features rather than sharp modes. This is a consequence of the neutron scattering selection rule $\Delta$S=0,$\pm$1 which prevents a single spinon from being created, instead they must be created in multiple pairs to conserve spin angular momentum.

A spinon continuum was recently observed in Herbertsmithite which is the best realization of the spin-\textbf{\textonehalf} kagome antiferromagnet. Diffuse excitations were found that are dispersionless at low energy transfers \cite{Han12}. The spectrum of Herbertsmithite has many similarities to \cacro, both show broad, hexagonal rings with no evidence for an energy gap. There are however several differences, firstly \cacro has two bands of excitations rather than the single band observed in Herbertsmithite, this might simply be a consequence of its bilayer structure and more complex Hamiltonian. Second and more significantly, although both compounds are based on kagome layers the sizes of the hexagonal rings are different. While Herbertsmithite shows strong signal along the fourth Brillouin zone boundary (red lines in Fig.~\ref{fig:INS_theory}a), the low energy band of \cacro exhibits smaller hexagonal rings at the first Brillouin zone boundary (black lines in figure~\ref{fig:INS_theory}a). Additional weaker features are observed along the dashed lines in figure~\ref{fig:INS_theory}a. 

Dominant scattering at the first Brillouin zone boundary suggests significant antiferromagnetic correlations with a distance that is twice the first neighbor \Cr-\Cr distance. In order to test this idea, the data at 0.25~meV were compared to the equal time structure factor for a collection of uncorrelated valence bonds on a triangular lattice with lattice parameter twice the first neighbor distance in the kagome plane  (figure~\ref{fig:INS_theory}b). Even though this model neglects correlations within the small Kagome triangles it approximately captures the observed intensity modulation.

To gain further insight into the properties of this compound we performed Pseudofermion Functional Renormalization Group (PFFRG) calculations \cite{Reu10,Reu11,Sut14,Reu14}. The PFFRG technique can compute the static susceptibility and accurately determine whether a specific Hamiltonian develops static long-range magnetic order. PFFRG calculations performed using the Hamiltonian of \cacro clearly show the absence of static magnetism verifying the experimental results. The calculated susceptibility at $Q=0$ can be compared to the measured DC susceptibility (Fig.~\ref{fig:INS_theory}c), both increase smoothly with decreasing temperature and show no sharp features that could indicate a phase transition. 

Figure~\ref{fig:INS_theory}d shows the PFFRG susceptibility as a function of wavevector within the kagome bilayer at $E=0$~meV and $T=0$~K, it reproduces the diffuse hexagonal-rings observed in the data at 0.25~meV (Fig.~\ref{fig:INS_theory}a). While the theoretical correlations are clearly short-range, their intensity varies being strongest near the vertices of the hexagon. This is where Bragg peaks would be observed in the case of 120$^\circ$ N\'eel order on a triangular lattice antiferromagnet with a lattice constant twice the \Cr-\Cr first neighbor distance confirming the presence of dominant antiferromagnetic correlations on these length scales in \cacro. Additional scattering along the dashed lines in Fig.\ref{fig:INS_theory}a that are not reproduced by the ground state PFFRG calculations might be observed because the neutron scattering data is collected at a finite energy (the high incoherent background at $E=0$~meV makes comparison of the static susceptibility difficult). Alternatively weak higher order terms in the Hamiltonian such as Dzyaloshinskii-Moriya or further neighbor interactions could perturb the spins giving weight to these features, although both susceptibility and heat capacity data suggest that such terms must be very weak.

Taken together, the experimental and theoretical results described here provide strong evidence that \cacro is a quantum spin liquid characterized by slow dynamical fluctuations, static magnetism and excitations that are diffuse and reminiscent of spinon continua. The remaining question is why the Hamiltonian of \cacro supports spin liquid behavior? This Hamiltonian has never been studied before and is much more complex than the models currently proposed for resonating valence bond states. Furthermore, with significantly stronger ferromagnetic interactions compared to the antiferromagnetic interactions and no anisotropy, the source of frustration is unclear. The clue comes from the inelastic neutron scattering data at low energy transfers and the PFFRG calculation which show that \cacro has short-range correlations similar to a triangular antiferromagnet with lattice parameter twice that of the first neighbor \Cr-\Cr distance.

Within each of the kagome layers that form the bilayer, the dominant ferromagnetic triangles couple the three spin-\textbf{\textonehalf} \Cr ions into effective spin-\nicefrac{3}{2} objects. The weaker antiferromagnetic interactions then couple these spin-\nicefrac{3}{2} objects into an effective triangular lattice (given by the dashed orange lines in Fig.~\ref{fig:structure_hamiltonian}d). The ground state of a spin-\nicefrac{3}{2} triangular antiferromagnet is well-established both theoretically and experimentally to have static long-range magnetic order where nearest neighbor spins point 120{\textdegree} with respect to each other, while the excitations are sharp and can be approximated by renormalized spin-wave theory. The additional source of frustration which prevents \cacro from developing long-range magnetic order and causes the spin-waves to fractionalize into spinons must therefore arise from the intrabilayer coupling $J0$. The most striking feature of the bilayer arrangement is that the ferromagnetic triangles in one layer are directly coupled to the antiferromagnetic triangles in the other layer (and vice versa), because of this neither the ferromagnetic triangles nor the antiferromagnetic triangles can realize their preferred spin arrangement (parallel and 120{\textdegree} order respectively) as illustrated figure~\ref{fig:structure_hamiltonian}e. Although the intrabilayer coupling is the weakest interaction in \cacro and furthermore is ferromagnetic, it plays the crucial role in destabilizing long-range magnetic order and giving rise to a resonating valence bond ground state \cite{And73}.

To conclude we have shown both experimentally and theoretically that \cacro is a new quantum spin liquid with persistent slow dynamics in the ground state and spinon excitations. This long-sought after state arises from a new and unexpected frustration mechanism where ferromagnetic triangles are directly coupled to antiferromagnetic triangles. The excitations are gapless and the correlations are similar to those of the triangular lattice antiferromagnet but with enhanced fluctuations strong enough to destroy both the static long-range order and the spin-wave excitations. This contrasts with the usual models for spin liquids which have competing antiferromagnetic interactions (e.g. kagome) or anisotropy in competition with the interactions (e.g. spin-ice). Thus quantum spin liquid behavior is not restricted to the simple models currently proposed but can be more widespread and exist in more complex structures than has previously been assumed.

\vspace{10pt}
\noindent\textbf{\large Methods}\\
\textbf{Sample preparation.} The powder samples were prepared from high purity powders of CaCO$_3$ (Alfa Aesar, 99.95\%) and Cr$_2$O$_3$ (Alfa Aesar, 99.97\%). The starting materials were mixed thoroughly in a 3:1 ratio and calcined in air at 1000\degree C for 24 hours. The single crystal was grown by the traveling-solvent floating-zone technique using an optical image furnace (Crystal Systems Corp., Japan). A cylindrical feed rod was prepared from the powder which was pressed hydrostatically up to 3000~bar in a cold isostatic press and sintered in air at 1010\degree C for 12 hours followed by rapid quenching to room temperature. A solvent with the composition 71.5~mol\% CaCO$_3$ and 28.5~mol\% Cr$_2$O$_3$ was prepared in the same way and about 0.5~g of it was attached to the tip of the feed rod to start the growth. A stable growth was achieved under an oxygen atmosphere of 2~bar and a growth rate of 1~mm/h. Two single crystalline pieces each about 15~mm in length and 6~mm in diameter were obtained by this process.\\
\textbf{Measurements.} The specific heat was measured down to 300~mK on a 0.91~mg single crystal using a relaxation technique at the Laboratory for Magnetic Measurements, Helmholtz Zentrum Berlin f\"ur Materialien und Energie, Germany. AC susceptibility was measured down to 50~mK on a 49~mg single crystal using a compensated coil pair system at the Dresden High Magnetic Field Laboratory, Helmholtz Zentrum Dresden Rossendorf, Germany. The \musr measurements took place on the LTF and GPS spectrometers at the Laboratory for Muon Spin Spectroscopy, Paul Scherrer Institute, Switzerland. For the LTF measurement, 200 mg of powder were spread on a Ag backing plate and mixed with a small drop of alcohol diluted GE Varnish for better thermal conductivity. The measurement on GPS was performed on a 300 mg powder sample in a thin Ag foil packet. Zero- and longitudinal-field measurements were performed for temperatures from 19~mK to 210~K. The data were analyzed using the musrfit software package \cite{Sut12}. The powder inelastic neutron scattering experiment was performed on the time-of-flight spectrometer TOFTOF at the Heinz Maier Leibnitz Zentrum, Munich, Germany at a temperature of 430~mK using 8.14~g of powder. An incident energy of 3.27~meV was used giving a resolution of 0.08~meV. The data was binned into steps of 0.02~{\AA} and 0.02~meV. The single crystal inelastic neutron scattering experiments were performed on the triple-axis spectrometers IN14 with the flat cone option at the Intitut Laue-Langevin, Grenoble, France and MACS II at the NIST Center for Neutron Research, Gaithersburg, USA. Additional data was taken on the indirect time-of-flight spectrometer OSIRIS at the ISIS facility, Didcot, UK. At IN14 two single crystals with masses of 0.97~g and 0.74~g were measured with scattering planes $(h0l)$ and $(hk0)$ respectively. The measurements took place at a temperature of 1.6~K, the final neutron energy was fixed to 4.06~meV giving a resolution of 0.12~meV. The data was binned into steps of 0.0085 along $[h,h,0]$, 0.013 along $[h,-h,0]$ and 0.06 along $[0,0,l]$ and the binned slices were then smoothed by a weighted average over 1.5 bins. For the MACS II experiment the two single crystals (total mass 1.71~g) were co-aligned with the $(hk0)$ plane horizontal and were measured at 90~mK. The final energy was fixed at 3.7~meV giving a resolution of 0.33~meV. The data was binned into steps of 0.035 along $[h,h,0]$ and 0.06 along $[h,-h,0]$. The same specimen was also used on OSIRIS at a temperature of 220~mK and a energy resolution of 0.025~meV. The data was binned into steps of 0.05~r.l.u. and 0.05~meV and smoothed with a hat function of width 2 bins.\\
\textbf{Functional renormalization group calculations.} In the PFFRG approach, the spin Hamiltonian is first rewritten in terms of Abrikosov pseudofermions as
\begin{equation}
S_i^\mu=\frac{1}{2}\sum_{\alpha\beta}f_{i\alpha}^\dagger \sigma_{\alpha\beta}^{\mu}f_{i\beta}\;,
\end{equation}
with $\alpha,\beta=\uparrow,\downarrow$. Here $f_{i\alpha}^{(\dagger)}$ denotes a pseudofermion annihilation (creation) operator of spin $\alpha$ at site $i$, and $\sigma_{\alpha\beta}^\mu$ are the Pauli matrices ($\mu=x,y,z$). The fermionic representation enables us to apply Wick's theorem, leading to a diagrammatic Feynman many-body expansion in the exchange couplings $J$. Since spin models are inherently strongly coupled quantum systems without a small parameter, diagrammatic summations need to be performed in infinite order in $J$ which is accomplished by the FRG approach. This technique introduces an artificial infrared frequency cutoff $\Lambda$ in the fermionic Green's function which suppresses the fermion propagation below the energy $\Lambda$. The FRG (for recent reviews, see e.g. \cite{Met12,Pla13}) then formulates an infinite hierarchy of coupled differential equations for the evolution of all $m$-particle vertex functions under the flow of $\Lambda$. In order to obtain a closed set of equations, three-particle vertices are neglected \cite{Reu10}. A crucial advantage of the PFFRG is that the diagrammatic summation includes vertex corrections between all two-particle interaction channels, i.e., it treats magnetic ordering and disordering tendencies on an equal footing. The central outcome of the PFFRG is the $Q$-space resolved magnetic susceptibility which directly follows from the fermionic two-particle vertex. Most importantly, the FRG parameter $\Lambda$ is proportional to the temperature $T$. Numerical benchmarks for an Ising model suggest that $T=\Lambda$ \cite{Goe12}, which has also been used for the comparison of PFFRG and experimental data in Fig.~\ref{fig:INS_theory}c. The evolution of the susceptibility as a function of $\Lambda$ (i.e. $T$) enables us to probe the system with respect to magnetic order: a smooth divergence-free flow without signatures of an instability hints at a non-magnetic ground state, while a sharp cusp and breakdown of the $\Lambda$-flow indicates the onset of magnetic order.

\bibliography{D:/Profile/isz/Bib/Cbalz}

\vspace{10pt}
{\small\noindent\textbf{Acknowledgments}\\
We thank Sandor Toth for his help with the SpinW program. We acknowledge the Helmholtz Gemeinschaft for funding via the Helmholtz Virtual Institute (Project No. HVI-521) and DFG through Research Training Group GRK 1621 and SFB 1143. We also acknowledge the support of the HLD-HZDR, a member of the European 
Magnetic Field Laboratory (EMFL). This work utilized facilities supported in part by the National Science Foundation under Agreement No. DMR-1508249.}

{\small\noindent\textbf{Author contributions}\\
CB performed or participated in all measurements and did the data analysis with help from the other authors. BL directed the project, participated in most measurements and wrote the manuscript with contributions from all authors. JR carried out the PFFRG calculations and provided theoretical insight. YS introduced the compound and made the powder while the crystals were grown by YS and ATMNI. HR carried out the specific heat measurements; RS and TH performed the AC susceptibility measurements and helped with the analysis; CB and HL helped with the \musr measurements with the analysis; GGS, EMW, TG, and JARR supported the INS measurements.}

{\small\noindent\textbf{Additional information}\\
Supplementary information is available in the online version of the paper. Correspondence and requests for materials should be addressed to CB.}

{\small\noindent\textbf{Competing financial interests}\\
The authors declare no competing financial interests.}



\section*{Supplementary material}

# Physical realization of a quantum spin liquid based on a novel frustration mechanism
## - Supplementary information -

Christian Balz,[1,2,*] Bella Lake,[1,2] Johannes Reuther,[1,3] Hubertus Luetkens,[4] Rico Schönemann,[5] Thomas Herrmannsdörfer,[5] Yogesh Singh,[6] A.T.M. Nazmul Islam,[1] Elisa M. Wheeler,[7] Jose A. Rodriguez-Rivera,[8,9] Tatiana Guidi,[10] Giovanna G. Simeoni,[11] Chris Baines,[4] and Hanjo Ryll[1]

[1]*Helmholtz-Zentrum Berlin für Materialien und Energie, 14109 Berlin, Germany*
[2]*Institut für Festkörperphysik, Technische Universität Berlin, 10623 Berlin, Germany*
[3]*Dahlem Center for Complex Quantum Systems and Fachbereich Physik, Freie Universität Berlin, 14195 Berlin, Germany*
[4]*Laboratory for Muon-Spin Spectroscopy, Paul Scherrer Institut, 5232 Villigen, Switzerland*
[5]*Hochfeld-Magnetlabor Dresden (HLD-EMFL), Helmholtz-Zentrum Dresden-Rossendorf, 01314 Dresden, Germany*
[6]*Indian Institute of Science Education and Research (IISER) Mohali, Mohali 140306, India*
[7]*Institut Laue-Langevin, 38042 Grenoble, France*
[8]*NIST Center for Neutron Research, National Institute of Standards and Technology, 20899 Gaithersburg, USA*
[9]*Department of Materials Science, University of Maryland, College Park, 20742 Maryland, USA*
[10]*ISIS Facility, STFC Rutherford Appleton Laboratory, Oxfordshire OX11 0QX, UK*
[11]*Heinz Maier-Leibnitz Zentrum, Technische Universitat München, 85748 Garching, Germany*

### Crystal structure and valence of the Cr ions

The crystal structure of $Ca_{10}Cr_7O_{28}$ has been determined at 183 K [1]. There are three inequivalent Cr ions, Cr1 and Cr2 occupy Wyckoff site 18b1, they form the kagome bilayer structure and are represented by black and gray spheres in figure **1a** of the main text, Cr3 occupies the 6a3 site and lie close to the centers of the hexagonal hole in the kagome bilayers (not shown). By performing a valance bond sum calculation the authors showed that Cr1 and Cr2 have valance 5+ while Cr3 has valance 6+. The $Cr^{5+}$ ions have spin-½ and are important for the magnetism while $Cr^{6+}$ is non-magnetic. Our own diffraction measurements on powder and single crystal samples using a combination of neutron and x-ray diffraction [2] agree with the published structure and furthermore show that the structure remains unchanged from 300 K down to 2 K. Our valence bond sum calculation gives valences of 4.88 for Cr1 and Cr2 and 5.95 for Cr3 in agreement with literature [1].

The absence of any phase transitions is further verified by the DC susceptibility data (Fig. **1a**) which reveals only smoothly varying features down to 2 K. The DC susceptibility also provides further evidence for the non-magnetic nature of the Cr3 ions. By fitting it to the Curie-Weiss law in the temperature range 50-250 K the effective moment per Cr ion yields $6/7 \cdot 1.74$ $\mu_B$ (where 1.74 $\mu_B$ is the effective moment for spin-1/2) suggesting that only 6 of 7 Cr ions are magnetic. Our magnetization measurements also support the non-magnetic nature of the Cr3 ion (Fig. **1b**). At a field of 12 T the magnetization saturates to 6/7 $\mu_B$ as expected if 1 in 7 Cr ions is non-magnetic which is indeed the number ratio of Cr3:(Cr1+Cr2). An earlier XANES measurement [3] of $Ca_{10}Cr_7O_{28}$ detected an average Cr valence of 5.3(1) which is consistent with the average valence 5.14 for a ratio of 6:1 among the 5+ and the 6+ Cr ions.

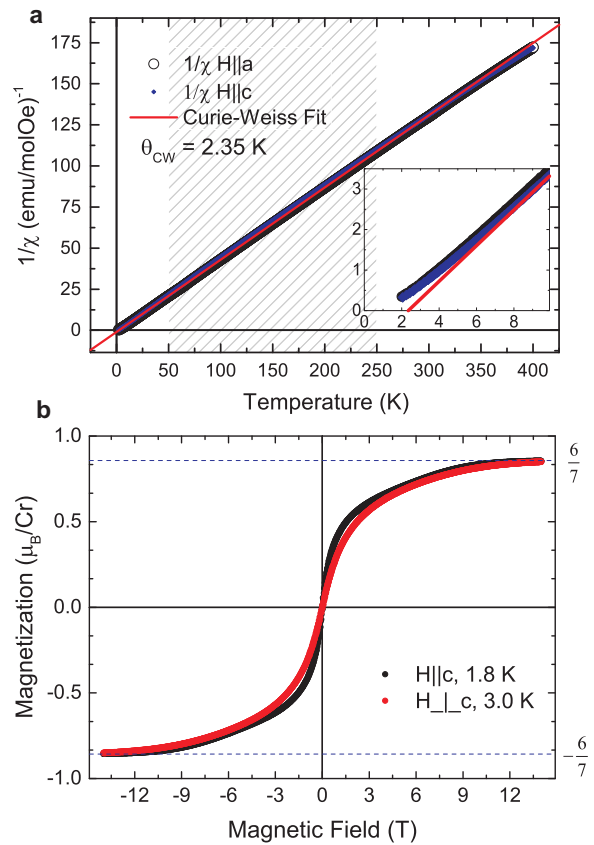

**Figure 1**. **Magnetization data. a**, Inverse susceptibility measured at 0.1 T. The red line is a fit to the Curie-Weiss law for $H||a$ in the temperature region indicated by the shading. The inset shows deviations from Curie-Weiss behavior at low temperatures. **b**, Magnetization at 1.8 K (3 K) with the magnetic field applied parallel to the $c$ axis (perpendicular to the $c$ axis).

## The Hamiltonian

A lot of information about the magnetic Hamiltonian could be extracted from the diffraction and bulk properties measurements. From the crystal structure refinement it is clear that the $CrO_4$ tetrahedra are always distorted leading to quenched orbital moments. This in turn yields spin-only moments and Heisenberg interactions. The DC susceptibility data confirms that the interactions are isotropic (Heisenberg) since no difference between the in-plane and out-of-plane field directions was observed (see Fig. **1a**). Furthermore, the small Curie-Weiss temperature which lies in the range $\pm 3$ K reveals that $Ca_{10}Cr_7O_{28}$ has a combination of ferromagnetic and antiferromagnetic interactions. The magnetization measurement reveals saturation at 12 T suggesting the magnetic excitations do not extend beyond 1.4 meV, therefore no individual exchange interactions is expected to be greater than 1.4 meV. Finally, in agreement with the specific heat (Fig. **2b** of the main text), there is no indication of an energy gap between the ground state and first excited state which excludes dimerization due to a single dominant antiferromagnetic interaction.

The most important result of the magnetization measurement is however that a magnetic field of 12 T is sufficient to overcome the interactions and force the spins to point in the direction of the external field. At this field we expect spin-wave excitations which can easily be measured using inelastic neutron scattering and fitted to linear spin-wave theory to extract the exchange interactions. Figure **2a** shows the inelastic neutron scattering data collected at 11 T. Gapped dispersive excitations are clearly visible along the $[h, -h, 0]$ direction in the plane of the kagome bilayers. The dispersions are due to the interactions between the ions within this plane. The two-dimensionality of the magnetic interactions is established in figure **4e** of the main text where the diffuse magnetic excitations measured in zero field in the $(h0l)$ plane at an energy transfer of 0.25 meV are shown. In contrast to the excitations in the $(hk0)$ (kagome) plane shown in figure **4b** of the main text no dispersion along the out-of-plane $l$-direction is visible. This indicates the absence of continuous magnetic interactions in this direction. The modulation of intensity along this direction is solely caused by the structure factor of the $J0$ bond.

The dispersive excitations at $H = 11$ T shown in figure **2a** were fitted to linear spin wave theory using the SpinW Matlab library [4] in the following way: The magnetic structure was assumed to ferromagnetic with all spins aligned along the field direction. The possible inequivalent nearest-neighbor exchange interactions were identified from the crystal structure (see Fig. **1a** of the main text). A Hamiltonian was then generated with random starting parameters which were fitted to the spin-wave spectra using a non-linear least square minimization routine. For this purpose the spin-spin correlation function was calculated from the Hamiltonian and compared to the neutron scattering cross section. For generating the random starting parameters the constraints obtained from the bulk properties as well as from the INS measurements (such as $|J_n| < 1.4$ meV) were used. Further we restricted ourselves to the seven shortest Cr-Cr distances in the unit cell of $Ca_{10}Cr_7O_{28}$. The best agreement to the data was achieved by the parameters given in table **1b** of the main text. The calculated spin-wave modes for these parameters are plotted over the INS data in figure **2a** and the simulated intensity distribution is shown in figure **2b**. These parameters lead to the kagome bilayer model where $J11$ and $J12$ are set to zero to satisfy the constrain of no continuous magnetic coupling along the $l$ direction. Magnetic models where $J0$ is zero while $J11$ and $J12$ are finite or even all three out-of-plane interactions ($J0$, $J11$ and $J12$) are zero could not produce fits of similar accuracy.

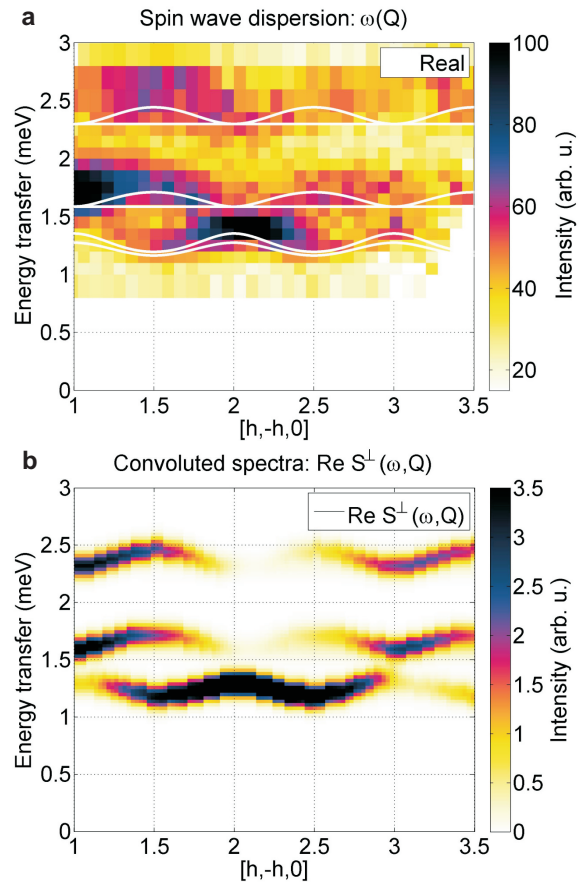

**Figure 2**. **Inelastic neutron scattering data under external magnetic field. a**, Inelastic neutron scattering along the $[h, -h, 0]$ axis at 11 T and 90 mK measured using the MACS spectrometer. The dispersion obtained from linear spin wave theory is overplotted. **b**, Intensity distribution calculated from linear spin wave theory using the best fit parameters convoluted with the instrumental resolution.

* christian.balz@helmholtz-berlin.de



\end{document}